\RequirePackage{color}

\documentclass[11pt,a4paper]{JHEP3} 
\usepackage{amssymb,amsmath}
\usepackage{euscript}
\usepackage{slashed}
\usepackage{amsfonts}
\usepackage{verbatim}
\usepackage[latin9]{inputenc}
\usepackage{amsthm}
\usepackage{esint}
\usepackage{color}
\def\be{\begin{eqnarray}}
 \def\ee{\end{eqnarray}}
 \def\0{\nonumber}
\def\tr{\rm tr}
\def\Tr{\rm Tr}

\def\e{\epsilon}

\def\EW{\EuScript{W}}
\def\EF{\EuScript{F}}\def\EA{\EuScript{A}}
\preprint{SISSA/03/2018/FISI\\ZTF-EP-18-01\\{\tt hep-th/1802.02968 } }

\title{Worldline quantization of field theory, effective actions and $L_\infty$
structure}

\author{ L.~Bonora$^{a}$, M.~Cvitan$^{b}$, P.~Dominis
Prester$^{c}$, S.~Giaccari$^{b}$, M.~Pauli\v{s}i\'c$^{c}$, T.~$\bf
\check{S}$temberga$^{b}$
\\\textit{${}^{a}$ International School for Advanced Studies (SISSA),\\Via
Bonomea 265, 34136 Trieste, Italy, and INFN, Sezione di
Trieste\\}%
\textit{${}^{b}$ Department of Physics, Faculty of Science, University
of Zagreb, \\ 
Bijeni\v{c}ka cesta 32, 10000 Zagreb, Croatia\\}%
\textit{${}^{c}$ Department of Physics, University of Rijeka,\\
Radmile Matej\v{c}i\'{c} 2, 51000 Rijeka, Croatia\\}%
E-mail: \email{bonora@sissa.it}, \email{mcvitan@phy.hr}, 
\email{pprester@phy.uniri.hr}, \email{sgiaccari@phy.hr},
\email{mateo.paulisic@phy.uniri.hr}, 
\email{tstember@phy.hr}}

\abstract{We formulate the worldline quantization  {(a.k.a. deformation quantization)} of a massive fermion model
coupled to external 
higher spin sources. We use the relations obtained in this way to show that its
regularized effective 
action is endowed with an $L_\infty$ symmetry. The same result holds also for a
massive scalar model. }


\begin{document}

\section{Introduction}
\label{sec:intro}
It is a widespread conviction, and arguments in favor of it are not lacking
\cite{Maldacena}, that, for a consistent quantum theory
of gravity and matter, an infinite number of fields is needed. This is so, of
course, in the case of (super)string theories, where infinite towers of
higher-spin excitations conspire to regulate the singular high-energy behavior
present in perturbatively quantized Einstein gravity. Other higher spin theories
exist in a four-dimensional and lower space-time,
see \cite{Vasiliev}. Very likely these are not the only possibilities. But then
a question arises: what are the requirements to be satisfied in order for these
theories to make sense? In particular, how can a high energy behavior like in
(super)string theories be guaranteed? In the latter this is tied to the short
distance behavior and has to do with the finite string size. So it is related to
the mild form of non-locality in string theory. In general, what is the right
amount of non-locality? All these are very general questions for which answers
are not yet available. For the time being we have to content ourselves with the
taxonomy of higher spin models.

Recently we have revisited and generalized a method based on effective actions
to determine the classical dynamics of higher spin fields,
\cite{BCLPS,BCDGLS,BCDGS}. The basic idea is to exploit the one-loop effective
actions of elementary free field theories coupled via conserved currents to
external higher spin  sources, in order to extract information about the
(classical) dynamics of the latter. We focused on massive scalar and Dirac
fermion models, but, no doubt, the same method can be applied to other
elementary fields. In the cited papers we computed the two-point correlators of
conserved currents, which allowed us to reconstruct the quadratic  effective
action for the higher spin fields coupled to the currents. We were able to show
that such effective actions  are built out of the Fronsdal differential
operators \cite{Fronsdal}, appropriate for those higher spin fields, in the
general non-local form discussed in \cite{FS}.

The method we used in \cite{BCLPS,BCDGLS,BCDGS} is the standard perturbative
approach 
based on Feynman diagrams.  This method is ultra-tested and very effective for
two-point
correlators. For instance, as we have seen in  \cite{BCDGS}, it preserves gauge
and 
diff-invariance (it respects the relevant Ward identities). We have no reason to
doubt that this will be the case also for higher order correlators, in
particular for the crucial three-point ones. But the burden to guess what the
gauge transformations beyond the lowest level are is left to us. In this regard
there exists an alternative quantization method which can come to our help, the
worldline quantization method\footnote{The literature on the worldline
quantization is large. Here we refer in particular to the calculation of
effective actions via the worldline quantization in relation to higher spin
theories,
\cite{Segal, bekaert}. The first elaboration of this method is probably in
\cite{Strassler}, to which  
many others followed, see for instance
\cite{Schmidt,DHoker,Bastianelli,Dai,Bonezzi}.}, 
which we wish to discuss in this paper. 

The worldline quantization of field theory is based on the Weyl quantization of
a particle in quantum mechanics, where the coordinates in the phase space are
replaced by position and momentum operator and observables are endowed with a
suitable operator ordering. In order to achieve second quantization one, roughly
speaking, replaces the field dependence on the position and the field
derivatives by the corresponding position and momentum operators, respectively,
and relies on the Weyl quantization for the latter. The effective action is then
defined. The important thing is that this
procedure comes with a bonus, the precise form of the gauge symmetry. This has a
remarkable consequence, as we will show in this paper: without doing explicit
calculations, it is possible to establish the symmetry of the full (not only the
local part of) effective action and demonstrate its $L_\infty$ symmetry. The 
latter is a symmetry that characterizes many (classical) field theories,
including closed string 
field theory (a good introduction to $L_\infty$ algebras and field theory is
\cite{HZ}). 

In section 2 we will carry out the worldline quantization of free Dirac fermions
coupled to external sources (the case of a scalar field has already been worked
out in  \cite{bekaert}) and derive heuristic rules, similar to the Feynman
diagrams, to compute amplitudes. 
In section 3 we will uncover the $L_\infty$ structure
of the corresponding effective action. Section 4 is devoted to a summary and
discussion
of our results.

\section{Worldline quantization of a fermion model}

\subsection{ Fermion linearly coupled to higher spin fields}

Let us consider a free fermion theory 
\be 
S_{0}= \int d^dx \, \overline \psi(i \gamma\!\cdot\! \partial-m)\psi ,\label{S0}
\ee
coupled to external sources. We second-quantize it using the Weyl quantization
method for a particle worldline.
The full action is expressed as an expectation value of operators as follows
\be
S= \langle \overline \psi | - \gamma \!\cdot\! (\widehat P- \widehat H) -m
|\psi\rangle \label{S}
\ee   
Here $\widehat P_\mu$ is the momentum operator whose symbol is the classical
momentum $p_\mu$. 
$\widehat H$ is an operator whose symbol is 
$h(x,p)$, where
\be
h^\mu(x,p)=\sum_{n=0}^\infty \frac 1{n!}\, 
h_{(s)}^{\mu\mu_1\ldots\mu_n} (x)\, p_{\mu_1}\ldots p_{\mu_n}\label{hmmm}
\ee 
$s=n+1$ is the spin and the tensors are assumed to be symmetric. We recall that
a quantum operator $\widehat O$ can be represented with a symbol $O(x,p)$
through 
the Weyl map
\be
\widehat O = \int  {d^d x}\,  {d^d y}  \frac{d^dk}{(2\pi)^d}
\frac{d^dp}{(2\pi)^d} \,  O(x,p)\, e^{i k \cdot (x-\widehat
X)-iy\cdot(p-\widehat
P)}\label{hatO}
\ee  
where $\widehat X$ is the position operator.

Next we insert this into the RHS of \eqref{S}, where we also insert  two
completenesses $\int
d^d x |x\rangle \langle x|$, and make the identification $\psi(x)= \langle
x|\psi\rangle$. Expressing $S$ in terms of symbols we find
\be
S&=& S_0  +  \int \frac {d^d q}{(2\pi)^d}\,  {d^d x}\, d^dz \, e^{i q\cdot z}\,
\overline \psi\! \left(x+\frac z2\right) \gamma\!\cdot\! h(x,q)\,
\psi\!\left(x-\frac
z2\right) \label{Ssymbol}\\
&=& S_0  + \sum_{n=0}^\infty \int  {d^d x}\, \frac{i^n}{n!}\,  \frac
{\partial}{\partial z^{\mu_1} }\ldots  \frac {\partial}{\partial z^{\mu_n}}
\overline \psi\! \left(x+\frac z2\right)  \gamma_\mu
h^{\mu\mu_1\ldots\mu_n}(x)\, \psi\!\left(x-\frac z2\right) \Big{\vert}_{z=0}
\nonumber \\
&=& S_0 + \sum_{s=1}^\infty \int  {d^d x}\, J^{(s)}_{\mu_1\ldots\mu_s}(x)\,
h_{(s)}^{\mu_1\ldots\mu_s}(x)
\nonumber
\ee 
We see that the symmetric tensor field $h^{\mu\mu_1\ldots\mu_n}$ is linearly
coupled to the HS (higher spin) current
\be
J^{(s)}_{\mu\mu_1\ldots\mu_{s-1}}(x) = \frac {i^{s-1}} {(s-1)!} \frac
{\partial}{\partial z^{(\mu_1}
}\ldots  \frac {\partial}{\partial z^{\mu_{s-1}}} \overline \psi \left(x+\frac
z2\right)  \gamma_{\mu)} \psi\left(x-\frac z2\right)
\Big{\vert}_{z=0}.\label{jmmm}
\ee
For instance, for $s=1$ and $s=2$ one obtains
\be
J^{(1)}_{\mu}&=& \overline \psi \gamma_\mu\psi\label{J1}\\
J^{(2)}_{\mu\mu_1}&=& \frac i2 \left( \partial_{(\mu_1} \overline \psi
\gamma_{\mu)}\psi-  \overline \psi
\gamma_{(\mu}\partial_{\mu_1)}\psi\right)\label{J2}
\ee
{{The HS currents are on-shell conserved in the free theory}} (\ref{S0})
{ \be \label{curcons}
\partial_\mu J_{(s)}^{\mu\mu_1\cdots\mu_{s-1}} = 0
\ee
which is a consequence of invariance of $S_0[\psi]$ on global (rigid)
transformations
\be \label{freeclaw}
\delta_n \psi(x) ={-\frac{(-i)^{n+1}}{n!}}\,
\varepsilon_{(n)}^{\mu_1\cdots\mu_n} \partial_{\mu_1} \ldots \partial_{\mu_n}
\psi(x)
\ee
We shall next show that for the full action} (\ref{Ssymbol}) { this extends to
the local symmetry. The consequence is that the currents are still conserved,
with the HS covariant derivative substituting ordinary derivative in}
(\ref{curcons}).

Notice that these currents are conserved even without symmetrizing $\mu$ with
the other indices. But in the sequel we will suppose that they are symmetric.

\subsection{Symmetries}

The action \eqref{S} is trivially invariant under the operation
\be
S= \langle \overline \psi | {\widehat O}  {\widehat O}^{-1} \widehat G{\widehat
O}{\widehat O}^{-1}|\psi\rangle 
\ee
where $\widehat G=  - \gamma \!\cdot\! (\widehat P- \widehat H) -m$. So it is
invariant under
\be
\widehat G \longrightarrow {\widehat O}^{-1} \widehat G{\widehat O},\quad\quad
|\psi\rangle \longrightarrow {\widehat O}^{-1}|\psi \rangle\label{G'}
\ee
Writing $\widehat O = e^{-i \widehat E}$ we easily find the infinitesimal
version.
\be
\delta |\psi \rangle = i \widehat E |\psi\rangle,\quad\quad \delta \langle
\overline \psi| =- i  \langle \overline \psi| \widehat E, \label{deltapsi}
\ee
and
\be 
\delta \widehat G= i[\widehat E\,,\widehat G] = i[ \gamma\!\cdot \!(\widehat P-
\widehat H)\,,\widehat E] = \gamma \!\cdot\! \delta \widehat H\label{deltaG}
\ee
Let the symbol of $\widehat E$ be $\varepsilon(x,p)$, then the symbol of $
[i\gamma\!\cdot \!\widehat P, \widehat E]$ is
\be 
\int {d^d y} \langle x-\frac y2| [i\gamma\!\cdot \!\widehat P, \widehat E]|
x+\frac y2\rangle \,e^{i y\cdot p}\label{PE}
\ee
An easy way to make this explicit is to use the fact that the symbol of the
product of two operators is given by the Moyal product of the symbols. Thus
\be
\mathrm{Symb}\big( [\gamma\!\cdot \!\widehat P, \widehat E]\big) &=&
 [\gamma\!\cdot\! p \stackrel{\ast}{,} \varepsilon(x,p)]
= \gamma\!\cdot\! p\, e^{-\frac i2 \stackrel{\rightarrow}{\partial_x}\cdot
\stackrel{ \leftarrow} {\partial_p}}\varepsilon(x,p) -
\varepsilon(x,p)\, e^{\frac i2 \stackrel{\leftarrow}{\partial_x}\cdot
\stackrel{\rightarrow} {\partial_p}} \gamma\!\cdot\! p
\0 \\
&=& -i \gamma\!\cdot\!\partial_x  \varepsilon(x,p)
\label{PE'}
\ee
Similarly
\be
\mathrm{Symb}\big([\widehat H^\mu,\widehat E]\big) =
 [h^\mu (x,p) \stackrel{\ast}{,} \varepsilon(x,p)]   \label{HE}
\ee
where $[a \stackrel{\ast}{,} b] \equiv a \ast b - b \ast a$. Therefore, in terms
of symbols,
{ \be
\delta_\varepsilon h^\mu(x,p) = \partial^\mu_x 
\varepsilon(x,p)-i [h^\mu(x,p) \stackrel{\ast}{,} \varepsilon(x,p)] 
\equiv {\cal D}^{\ast\mu}_x  \varepsilon(x,p)
\label{deltahxp}
\ee}
where we introduced the covariant derivative defined by
\be
{\cal D}^{\ast\mu}_x = \partial_x^\mu - i  [h^\mu(x,p) \stackrel{\ast}{,}
\quad] 
\ee 

This will be referred to hereafter as HS transformation,
and the corresponding symmetry HS symmetry.

The transformations of $\psi$ are somewhat different. They can also be expressed
as Moyal product of symbols
\be
\delta_{\varepsilon} \tilde \psi(x,p) = i \varepsilon(x,p) \ast \tilde \psi
(x,p)\label{deltapsi1}
\ee
provided we use the partial Fourier transform
\be
\tilde \psi (x,p)= \int  {d^d y} \, \psi\left(x-\frac y2\right) e^{iy\cdot
p}.\label{tildepsi}
\ee
and finally we antitransform back the result. Alternatively we  can proceed as
follows. We compute
\be
 \langle x| \widehat E| \psi\rangle&=&  \int \frac {d^dk}{(2\pi)^d} \frac
{d^dp}{(2\pi)^d} d^dx' d^dy' \,\,\varepsilon(x',p) 
 \,\langle x|  e^{i k \cdot (x'-\widehat X)-iy'\cdot(p-\widehat P)}|\psi\rangle
\label{xEpsi}\\
&=& \int \frac {d^d k}{(2\pi)^d} \, \frac {d^d p}{(2\pi)^d} d^dx' d^dy' \,\,
\varepsilon(x',p')\, e^{ik\cdot (x'-x)-iy'\cdot p}  
\langle x| e^{i y' \widehat P} |\psi \rangle e^{-\frac i2 y'\cdot k}\0
\ee
Next we insert a momentum completeness $\int d^d q |q\rangle\langle q|$ to
evaluate $\langle x| e^{i y' \widehat P} |\psi \rangle$ and 
subsequently a coordinate completeness to evaluate $\langle q|\psi \rangle$
using the standard relation $\langle x|p\rangle = e^{ip\cdot x}$. 
Then we produce two delta functions by integrating over $k$ and $q$. In this way
we get rid of two coordinate integrations. Finally we arrive at
\be
\delta_\varepsilon \psi(x) &=& i \langle x| \widehat E|\psi\rangle
=   i \int \frac {d^d p}{(2\pi)^d}\, d^dz \,
\varepsilon\!\left(x+ \frac z2 ,p\right) e^{-i p\cdot z }\, \psi(x+z)
\label{xEpsi1}\\
&=& i \sum_{n=0}^\infty  \int \frac {d^d p}{(2\pi)^d}\, d^dz \, \frac {e^{-i
p\cdot
z }}{n!} (-i \partial_z)^n\cdot \left( \varepsilon_{(n)}\!\! 
\left( x+\frac z2\right) \psi(x+z)\right)\0\\
&=&  \sum_{n=0}^\infty \frac i{n!}  (-i \partial_z)^n\cdot \left(
\varepsilon_{(n)}\!\! 
\left( x+\frac z2\right) \psi(x+z)\right)\Big{\vert}_{z=0}\0\\
&=& i \varepsilon_{(0)}\!(x)\, \psi(x) + \varepsilon_{(1)}^\mu\!(x)\,
\partial_\mu
\psi(x) + \frac 12\partial_\mu \varepsilon_{(1)}^\mu\!(x)\, \psi(x)
\0 \\
&& -
\frac i2 \left( \varepsilon_{(2)}^{\mu\nu}\, \partial_\mu \partial_\nu \psi +
\partial_\mu \varepsilon_{(2)}^{\mu\nu}\, \partial_\nu \psi
 +\frac 14 
\partial_\mu \partial_\nu \varepsilon_{(2)}^{\mu\nu} \,\psi\right)\!\!(x) +
\ldots
\0
\ee
where a dot denote the contraction of upper and lower indices.
The first method leads to the same result.

{ Now we want to understand the conservation law ensuing from the
HS symmetry of the interacting 
classical action} (\ref{Ssymbol}) 
{ \be \0
0 = \delta_\varepsilon S[\psi,h] = \int d^dx \left( \frac{\delta
S}{\delta\psi(x)}\, \delta_\varepsilon \psi(x)
 + \delta_\varepsilon \overline{\psi}(x) \frac{\delta
S}{\delta\overline{\psi}(x)}
 + \int d^dp\, \frac{\delta S}{\delta h^\mu(x,p)}\, \delta_\varepsilon
h^\mu(x,p) \right)
\ee
Now we evaluate this expression on the classical solution, in which case the
first two terms vanish (remember that $h$ is the background field). We are left
with 
\be
0 = \int d^dx \int d^dp\, J_\mu(x,p)\, \delta_\varepsilon h^\mu(x,p) \qquad\quad
\mathrm{(on-shell)}
\ee
where
\be
J_\mu(x,p) \equiv \int \frac{d^dz}{(2\pi)^d}\, e^{i p \cdot z}\, \overline
\psi\! \left(x+\frac z2\right) \gamma_\mu\, 
\psi\!\left(x-\frac z2\right)
\ee
Using} (\ref{deltahxp}){, partially integrating and using the
following property of the Moyal product
\be \label{mypint}
\int d^dx \int d^dp\, a(x,p) [b(x,p) \stackrel{\ast}{,} c(x,p)] = \int d^dx \int
d^dp\, [a(x,p) \stackrel{\ast}{,} b(x,p)]\, c(x,p)
\ee
we obtain
\be
0 = \int d^dx \int d^dp\, \varepsilon(x,p)\, \mathcal{D}^{\ast\mu}_x J_\mu(x,p)
\qquad\quad \mathrm{(on-shell)}
\ee
From this follows the conservation law in the classical interacting theory
\be
\mathcal{D}^{\ast\mu}_x J_\mu(x,p) = 0 \qquad\qquad \mathrm{(on-shell)}
\ee
It is not hard to shaw that for $h^\mu(x,p) = 0$ this becomes equivalent to}
(\ref{freeclaw}).

Using the $\ast$-Jacobi identity (it holds also for the Moyal product, because
it is
associative) one can easily get
\be 
\left(\delta_{\varepsilon_2 }\delta_{\varepsilon_1 }- \delta_{\varepsilon_1
}\delta_{\varepsilon_2}\right) h^\mu(x,p) &=&  i\left( \partial_x
[{\varepsilon_1
}\stackrel{\ast}{,}{\varepsilon_2 }](x,p) -i [h^\mu(x,p)\stackrel{\ast}{,} 
[{\varepsilon_1 }\stackrel{\ast}{,}{\varepsilon_2 }](x,p) ]] \right)
\0 \\
&=&   i\, {\cal D}^{\ast\mu}_x [{\varepsilon_1}\stackrel{\ast}{,}{\varepsilon_2
}](x,p)
\label{e1e2}
\ee
We see that the HS $\varepsilon$-transform is of the Lie algebra type.

\subsection{Perturbative expansion of the effective action}

In this subsection we work out (heuristic) rules, similar to the Feynman ones, 
to compute $n$-point amplitudes in the above fermion model. The purpose is to
reproduce formulas similar to those
of \cite{bekaert} for the scalar case. We would like to point out, however, that
this is not strictly necessary: the good old Feynman rules are anyhow a valid
alternative.

We start from the representation of the effective action as trace-logarithm of a
differential operator:
\be
W[h] =N\, \Tr [\ln \widehat G] \label{G}
\ee
and use a well-known mathematical formula to regularize it
\be
W_{reg}[h,\epsilon]= -N \int_\epsilon^\infty  \frac {dt}t\, {\Tr}\! \left[ e^{-t
\widehat G}\right]
\label{Wreg}
\ee
where $\e$ is an infrared regulator.
The crucial factor is therefore  
\be
K[g|t] \equiv {\Tr}\! \left[e^{-t \widehat G}\right] = {\Tr}\! \left[ e^{t(  
\gamma
\cdot (\widehat P-\widehat H) +m)}\right],\label{Kgt}
\ee
known as the heat kernel,
where $g$ is the symbol of $\widehat G$. The trace $\Tr$ includes both an
integration
over the momenta and $\tr$, the trace over the gamma matrices,
\be
K[g|t] = e^{mt} \int  \frac {d^dp}{(2\pi)^d} \, {\tr} \langle p | e^{t \gamma
\cdot
(\widehat P-\widehat H)}|p\rangle\label{Kgt1}
\ee
Next we expand
\be
e^{ t  \gamma \cdot (\widehat P-\widehat H)}&=& e^{ t\, \gamma \cdot \widehat
P}\sum_{n=0}^\infty {(-1)^n}\int_0^t d\tau_1 \int_0^{\tau_1}
d\tau_2 \ldots
\int_0^{\tau_{n-1} }d\tau_n \, \gamma \!\cdot\! \widehat H(\tau_1)\,  \gamma
\!\cdot\!  \widehat H(\tau_2)\ldots   \gamma \!\cdot\! \widehat H(\tau_n)\0\\
\ee
where $ \gamma \! \cdot\! \widehat H(\tau)=  e^{ -\tau\,  \gamma \cdot \widehat
P}\gamma \!\cdot\! \widehat H\, e^{ \tau\,  \gamma \cdot \widehat P}$. 

We have
\be
\langle p|\gamma \! \cdot\! \widehat H(\tau)|q\rangle = e^{ -\tau\,  \gamma
\cdot p}\langle p|\gamma \!\cdot\! \widehat H|q\rangle\,  e^{ \tau\,  \gamma
\cdot
q}\label{pHq}
\ee
Using a formula analogous to \eqref{xEpsi} for $\widehat H$ and inserting
completenesses one finds
\be
\langle p|\gamma \!\cdot\! \widehat H|q\rangle 
&=& \int d^dx \int d^d y  \frac {d^dk}{(2\pi)^d}  \frac {d^dp'}{(2\pi)^d} \,
\gamma \!\cdot\! h(x,p') \langle p|  e^{i k \cdot ({x}-\widehat
X)-i{y}\cdot(p'-\widehat P)}|q\rangle \label{pHq1}\\
&=&  \int d^dx\, \gamma\!\cdot\! h(x, \partial_u) e^
{{i(q-p)}\cdot x+ u
\cdot\frac {p+q}2} \Big{\vert}_{u=0}\0
\ee
Therefore 
\be 
\langle p|\gamma \! \cdot\! \widehat H(\tau)|q\rangle =  \int d^dx\, e^{ -\tau\,
 \gamma \cdot p}\,\gamma\!\cdot\!  h(x, \partial_u)\, e^{ \tau\,  \gamma \cdot
q}\, e^ {{i(q-p)}\cdot x+ u \cdot\frac {p+q}2} \Big{\vert}_{u=0}
\ee

Using this we can write
\be
\Tr\left[ e^{-t \widehat G}\right]
&=& e^{mt} \sum_{n=0}^\infty {(-1)^n} \int \prod_{i=1}^n
\,{\frac
{d^dp_i}{(2\pi)^d}}  \int_0^t d\tau_1 \int_0^{\tau_1} d\tau_2 \ldots
\int_0^{\tau_{n-1} }d\tau_n \,\0\\
&\times& {\tr}\! \left({e^{t\,\gamma\cdot p_n}}\langle p_n|
\gamma \!\cdot\! \widehat H(\tau_1)|p_1\rangle
\langle p_1|  \gamma \!\cdot\!  \widehat H(\tau_2)|p_2\rangle\ldots   \langle
p_{n-1} |\gamma \!\cdot\! \widehat H(\tau_n)|p_n\rangle\right)\0\\
&=& e^{mt} \sum_{n=0}^\infty {(-1)^n}\int \prod_{i=1}^n
{d^dx_i} \frac {d^dp_i}{(2\pi)^d} 
\int_0^t d\tau_1 \int_0^{\tau_1} d\tau_2 \ldots \int_0^{\tau_{n-1} }d\tau_n \0\\
&\times& {\tr}\! \left( e^{(t-\tau_1)\,\gamma\cdot p_n} \gamma^{\mu_1}
e^{(\tau_1-\tau_2)\,\gamma\cdot p_1}\gamma^{\mu_2}
\ldots\gamma^{\mu_{n-1}}e^{(\tau_{n-1}-\tau_n)\,\gamma\cdot
p_{n-1}}\gamma^{\mu_n}e^{\tau_n \gamma \cdot p_n} \right)\0\\
&\times & \prod_{j=1}^n e^{i p_j\cdot {\left(x_j-x_{j+1} - i
\frac {u_{j+1}+u_j}
2\right)}  }\,\,
h_{\mu_1}\left(x_1,\stackrel{\leftarrow}{\partial_{u_1}}\right)\ldots 
h_{\mu_n}\left(x_n,\stackrel{\leftarrow}{\partial_{u_n}}\right)\Big{\vert}_{
u_j=0}\label{TretG}
\ee
where $x_{n+1}=x_1$.
Now we can factor out in $K[g,t]$ the terms $
h_{\mu_1}\left(x_1,\stackrel{\leftarrow}{\partial_{u_1}}\right)\ldots 
h_{\mu_n}\left(x_n,\stackrel{\leftarrow}{\partial_{u_n}}\right)$, and write
\be
{K[g|t]} =\sum_{n=0}^\infty  \langle\langle K^{(n)\mu\ldots
\mu}(t) |
h_\mu^{\otimes n}\rangle\rangle\label{Kgt/hn}
\ee
where the double brackets means integration of the $x_i$ and derivation with
respect to the $u_i$. In turn $K^{(n)\mu\ldots \mu}(t) $ can be written more
explicitly as 
\be
K^{\mu_1\ldots \mu_n}(x_1,u_1,\ldots,x_n,u_n|t)= e^{tm}  {\int
\prod_{j=1}^n\frac{ d^dp_j}{(2\pi)^d}} e^{i
p_j\cdot {\left(x_j-x_{j+1} - i \frac {u_{j+1}+u_j}
2\right)} }\widetilde
K^{\mu_1\ldots\mu_n}(p_1,\ldots,p_n|t) \qquad \label{Kmmm}
\ee
where
\be
&&\widetilde K^{\mu_1\ldots\mu_n}(p_1,\ldots,p_n|t)= \frac
{{(-1)^n}}n \int_0^t d\tau_1
\int_0^{\tau_1} d\tau_2 \ldots \int_0^{\tau_{n-1} }d\tau_n\label{Kmmmppp}\\
&&\times \,\, {\tr} \Big{(} \gamma^{\mu_1} e^{(\tau_1-\tau_2)\,\gamma\cdot
p_1}\gamma^{\mu_2}
\ldots\gamma^{\mu_{n-1}}e^{(\tau_{n-1}-\tau_n)\,\gamma\cdot
p_{n-1}}\gamma^{\mu_n}
e^{(\tau_n-\tau_1)\,\gamma\cdot p_n} e^{t \gamma \cdot p_n}\0\\
&&\quad +\,  \gamma^{\mu_2} e^{(\tau_1-\tau_2)\,\gamma\cdot p_2}\gamma^{\mu_3}
\ldots\gamma^{\mu_{n}}e^{(\tau_{n-1}-\tau_n)\,\gamma\cdot p_{n}}\gamma^{\mu_1}
e^{(\tau_n-\tau_1)\,\gamma\cdot p_1} e^{t \gamma \cdot p_1}\0\\
&&\qquad \vdots \0\\
&&\quad+\,  \gamma^{\mu_n} e^{(\tau_1-\tau_2)\,\gamma\cdot p_n}\gamma^{\mu_1}
\ldots\gamma^{\mu_{n-2}}e^{(\tau_{n-1}-\tau_n)\,\gamma\cdot
p_{n-2}}\gamma^{\mu_{n-1}}
e^{(\tau_n-\tau_1)\,\gamma\cdot p_{n-1}} e^{t \gamma \cdot p_{n-1}}\Big{)}\0
\ee

Now, the nested integral can be rewritten in the following way
\be
\int_0^t d\tau_1 \int_0^{\tau_1} d\tau_2 \ldots \int_0^{\tau_{n-1} }d\tau_n&=&
\int_0^t d\sigma_1 \int_0^{t-\sigma_1}d\sigma_2
\int_0^{t-\sigma_1-\sigma_2}d\sigma_3\ldots \int_0^{t-\sigma_1-\ldots
-\sigma_{n-1}} d\sigma_n\0\\
&=& \int_0^\infty  d\sigma_1\int_0^\infty  d\sigma_2 \ldots \int_0^\infty 
d\sigma_n\,  \theta (t-\sigma_1 -\ldots -\sigma_{n}) \label{tausigma1}
\ee
where $\sigma_i=\tau_{i-1}-\tau_i$, with $\tau_0=t$. Notice that defining
$\sigma_0= t-\sigma_1-\ldots -\sigma_n$ we can identify $\sigma_0=\tau_n$.

Next one uses the following representation of the Heaviside function
\be
\theta(t) = \lim_{\epsilon\to 0^+}\int_{-\infty}^{\infty} \frac{d\omega}{2\pi
i}\,
\frac {e^{i\omega t}}{\omega -i\epsilon} =  \lim_{\epsilon\to
0^+}\int_{-\infty}^{\infty} \frac{d\omega}{2\pi }\,  {e^{i\omega
t}}\int_0^\infty
d\sigma_0\, e^{-i \sigma_0 (\omega - i \epsilon)}
\label{thetat}
\ee
The $\omega$  integration has to be understood as a contour integration. { Using
this 
in} (\ref{tausigma1}) we obtain
\be
\int_0^t d\tau_1 \int_0^{\tau_1} d\tau_2 \ldots \int_0^{\tau_{n-1} }d\tau_n=
\int_{-\infty}^{\infty} \frac{d\omega}{2\pi }\, {e^{i\omega t}} \int_0^\infty
d\sigma_0 \int_0^\infty d\sigma_1\ldots \int_0^\infty  d\sigma_n\,
e^{-i(\sigma_0+
\ldots + \sigma_n) (\omega -i\epsilon)}\0\\
\label{thetat1}
\ee

Replacing this inside \eqref{Kmmmppp} we get 
\be
&&\widetilde K^{\mu_1\ldots\mu_n}(p_1,\ldots,p_n|t)=\frac
{{(-1)^n}}n\int_{-\infty}^{\infty} \frac{d\omega}{2\pi }\,
{e^{i\omega t}}   \int_0^\infty 
d\sigma_0\int_0^\infty d\sigma_1\ldots \int_0^\infty  d\sigma_n\,
\label{Kmmmppp1}\\
&\times& {\tr} \Big{[} \gamma^{\mu_1} e^{\sigma_2 (\gamma \cdot p_1 -i\omega')}
\gamma^{\mu_2} \ldots \gamma^{\mu_{n-1}} e^{\sigma_n (\gamma \cdot p_{n-1}
-i\omega')} \gamma^{\mu_n}  e^{(\sigma_0+\sigma_1) (\gamma \cdot
p_{n}-i\omega')}  \0\\
&& +\, \gamma^{\mu_2} e^{\sigma_2 (\gamma \cdot p_2 -i\omega')} \gamma^{\mu_3}
\ldots \gamma^{\mu_{n}} e^{\sigma_n (\gamma \cdot p_{n} -i\omega')}
\gamma^{\mu_1}  e^{(\sigma_0+\sigma_1) (\gamma \cdot p_1-i\omega')}  \0\\
&& \quad \vdots\0\\
&& +\, \gamma^{\mu_n} e^{\sigma_2 (\gamma \cdot p_n -i{\omega'}
)} \gamma^{\mu_1}
\ldots \gamma^{\mu_{n-2}} e^{\sigma_n (\gamma \cdot p_{n-2}
-i{\omega'} 
)}\gamma^{\mu_{n-1}}  e^{(\sigma_0+\sigma_1) (\gamma \cdot
p_{n-1}-i{\omega'}  )}
\Big{]}\0
\ee
where $\omega'= \omega- i \epsilon$. 
$\epsilon$ in the exponents allows us to perform the integrals\footnote{This is
evident
with the Majorana representation of the gamma matrices, because in such a case
the term $\gamma\!\cdot\!p$ in
the exponent is purely imaginary, the  gamma matrices being imaginary. This term
therefore gives rise to oscillatory
contributions, much like the $i\omega$ term.}, the result being  
\be
\widetilde K^{\mu_1\ldots\mu_n}(p_1,\ldots,p_n|t)&=&\frac
{{(-1)^n}}n
\int_{-\infty}^{\infty} \frac{d\omega}{2\pi }\, {e^{i\omega t}}
\label{Kmmmppp2}\\
&\times& {\tr} \Big{[} \gamma^{\mu_1} \frac {{-1}}{\slashed{ p}_1
-i\omega'}
\gamma^{\mu_2} \ldots \gamma^{\mu_{n-1}}\frac {{-1}}{
\slashed{p}_{n-1}-i\omega'}
\gamma^{\mu_n} \frac {1}{ (\slashed{ p}_{n}-i\omega')^2} \0\\
&& +\, \gamma^{\mu_2} \frac {{-1}}{ \slashed{ p}_2 -i\omega'}
\gamma^{\mu_3} \ldots
\gamma^{\mu_{n}} \frac {{-1}}{ \slashed{ p}_{n} -i\omega'}
\gamma^{\mu_1} \frac
1{(\slashed{ p}_1-i\omega' )^2 }\0\\
 && \quad \vdots \0\\
&& +\, \gamma^{\mu_n} \frac {{-1}}{ \slashed{ p}_n -i\omega'}
\gamma^{\mu_1} \ldots
\gamma^{\mu_{n-2}} \frac {{-1}}{ \slashed{p}_{n-2}
-i\omega'}\gamma^{\mu_{n-1}} \frac
1{ (\slashed{p}_{n-1}-i\omega')^2 } \Big{]}\0
\ee
We remark that  $\frac 1{(\slashed{p}-i\omega')^2}=\frac {\partial}{\partial
(i\omega)} \frac 1 {\slashed{p}-i\omega'}$. This allows us, via integration by
parts, to simplify \eqref{Kmmmppp2}
\be
\widetilde K^{\mu_1\ldots\mu_n}(p_1,\ldots,p_n|t)\!\!&=& \!\!\frac {t}n
\int_{-\infty}^{\infty} \frac{d\omega}{{2\pi }}\, e^{i\omega t}
\, {\tr} \Big{[}
\gamma^{\mu_1} \frac 1{\slashed{ p}_1\! - i\omega'} \gamma^{\mu_2} \ldots
\frac 1{ \slashed{p}_{n-1}\! - i\omega'} \gamma^{\mu_n} \frac 1{
\slashed{ p}_{n}\! - i\omega'}\Big{]} \qquad  
\label{Kmmmppp3}
\ee

We can also include the factor $e^{tm}$ in \eqref{Kmmm} in a new kernel
$\widetilde K^{\mu_1\ldots\mu_n}(p_1,\ldots,p_n|m,t)$
which has the same form as $\widetilde K^{\mu_1\ldots\mu_n}(p_1,\ldots,p_n|t)$
with all the $\slashed p_i$ replaced by $ \slashed p_i +m$:
\be
K^{\mu_1\ldots \mu_n}(x_1,u_1,\ldots,x_n,u_n|t)=  \prod_{j=1}^n e^{i p_j\cdot
{\left(x_j-x_{j+1} - i \frac {u_{j+1}+u_j}
2\right)}  }\widetilde
K^{\mu_1\ldots\mu_n}(p_1,\ldots,p_n|m,t) \qquad \label{Kmmmt}
\ee
\be
&&\!\!\!\!\!\!\!\!\!\!\!\!\widetilde
K^{\mu_1\ldots\mu_n}(p_1,\ldots,p_n|m,t) = \frac {{(-1)^n}}n
\int_{-\infty}^{\infty}
\frac{d\omega}{2\pi }\, e^{i\omega t} \label{Kmmmppp4}\\
&\times& {\tr} \Big{[} \gamma^{\mu_1} \frac  {{-1}}{\slashed{
p}_1 +m-i\omega'}
\gamma^{\mu_2} \ldots \gamma^{\mu_{n-1}}\frac  {{-1}}{
\slashed{p}_{n-1}+m-i\omega'}
\gamma^{\mu_n} \frac 1{ (\slashed{ p}_{n}+m-i\omega')^2} \0\\
&& +\, \gamma^{\mu_2} \frac  {{-1}}{ \slashed{ p}_2 +m-i\omega'}
\gamma^{\mu_3}
\ldots \gamma^{\mu_{n}} \frac  {{-1}}{ \slashed{ p}_{n}+m
-i\omega'} \gamma^{\mu_1}
\frac 1{(\slashed{ p}_1+m -i\omega')^2 } \0\\
 &&\quad \vdots \0\\
&& +\, \gamma^{\mu_n} \frac  {{-1}}{ \slashed{ p}_n +m-i\omega'}
\gamma^{\mu_1}
\ldots \gamma^{\mu_{n-2}} \frac  {{-1}}{ \slashed{p}_{n-2}+m
-i\omega'}\gamma^{\mu_{n-1}} \frac 1{ (\slashed{p}_{n-1}+m-i\omega')^2} 
\Big{]}\0\\ 
&=& { \frac tn} \int_{-\infty}^{\infty} \frac{d\omega}{2\pi }\,
e^{i\omega t}
\,{\tr} \Big{[} \gamma^{\mu_1} \frac 1{\slashed{ p}_1+m -i\omega'}
\gamma^{\mu_2}
\ldots \frac 1{ \slashed{p}_{n-1}+m-i\omega'} \gamma^{\mu_n}
\frac 1{ \slashed{ p}_{n}+m-i\omega'}\Big{]}  \0
\ee

Integrating further as in the scalar model case, \cite{bekaert}, is not possible
at this stage
because of the gamma matrices. One has to proceed first to evaluate the trace
over the latter.

Using \eqref{TretG} we can write the regularized effective action as
\be
W_{reg}[h,\epsilon] &=&-N \int_\epsilon ^\infty \frac {dt}t \,  e^{mt}
\sum_{n=0}^\infty \int \prod_{i=1}^n d^dx_i\, \frac {d^dp_i}{(2\pi)^d}  \int_0^t
d\tau_1 \int_0^{\tau_1} d\tau_2 \ldots \int_0^{\tau_{n-1} }d\tau_n \qquad
\0 \\
&& \times\, {\tr} \left( e^{(t-\tau_1)\cdot p_n} \gamma^{\mu_1}
e^{(\tau_1-\tau_2)\cdot p_1}\gamma^{\mu_2}
\ldots\gamma^{\mu_{n-1}}e^{(\tau_{n-1}-\tau_n)\cdot
p_{n-1}}\gamma^{\mu_n}e^{\tau_n \gamma \cdot p_n} \right)\0\\
&& \times\, \prod_{j=1}^n e^{i p_j\cdot {\left(x_j -x_{j+1}
\right)}  }\,\,
h_{\mu_1}\left(x_1,\frac {p_1+p_n}{2}\right)\ldots  h_{\mu_n}\left(x_n, \frac
{p_{n-1}+p_n}2\right)\0\\
&=& -N \int_\epsilon ^\infty dt \, \sum_{n=0}^\infty { \frac
1n}\int \prod_{i=1}^n d^dx_i\,
\frac {d^dp_i}{(2\pi)^d} \int_{-\infty}^{\infty} \frac{d\omega}{2\pi }\,
e^{i\omega t}\0\\
&& \times\,{\tr} \Big{[} \gamma^{\mu_1} \frac 1{\slashed{ p}_1+m -i\omega'}
\gamma^{\mu_2} \ldots \gamma^{\mu_{n-1}}\frac 1{ \slashed{p}_{n-1}+m-i\omega'}
\gamma^{\mu_n} \frac 1{ \slashed{ p}_{n}+m-i\omega'}\Big{]}  \0\\
&& \times\, \prod_{j=1}^n e^{i p_j\cdot  {\left(x_j -x_{j+1}
\right)} }\,\,
h_{\mu_1}\!\!\left(x_1,\frac {p_1+p_n}{2}\right)\ldots  h_{\mu_n}\!\!\left(x_n,
\frac
{p_{n-1}+p_n}2\right)
\label{Wreghm}
\ee

\subsection{Ward identities and generalized EoM}
\label{ssec:gEoM}

The general formula for the effective action is
\be
W[h] = \sum_{n=1}^\infty\, \frac 1{n!}\int \prod_{i=1}^n d^dx_i\,
\frac {d^dp_i}{(2\pi)^d}\,  \EW_{\mu_1,\ldots, \mu_n}^{(n)}(x_1,p_1,\ldots, x_n,
 p_n,\epsilon)\, h^{\mu_1}(x_1,p_1) \ldots  h^{\mu_n}(x_n,p_n)\0\\
 \label{EW}
\ee 
where we have discarded the constant 0-point contribution, as we will do
hereafter.
The effective action can be calculated by various methods, of which
\eqref{Wreghm} is a particular example.
In the latter case the amplitudes are given by
\be
\EW_{\mu_1,\ldots, \mu_n}^{(n)}(x_1,p_1,\!\!&\ldots&\!\!, x_n, 
p_n,\epsilon)= -N{ \frac {n!}n } \int_\epsilon ^\infty dt \, \int
\prod_{i=1}^n \frac
{d^dq_i}{(2\pi)^d} \int_{-\infty}^{\infty} \frac{d\omega}{2\pi }\, e^{i\omega t}
\0\\
&\times&\! {\tr} \Big{[} \gamma^{\mu_1} \frac 1{\slashed{ q}_1+m -i\omega'}
\gamma^{\mu_2} \ldots \gamma^{\mu_{n-1}}\frac 1{ \slashed{q}_{n-1}+m-i\omega'}
\gamma^{\mu_n} \frac 1{ \slashed{ q}_{n}+m-i\omega'}\Big{]}  \0\\
&\times&\! \prod_{j=1}^n e^{i q_j\cdot {\left(x_j 
-x_{j+1}\right) } }\,\,
\delta\!\left(p_1-\frac {q_1+q_n}{2}\right)\ldots   \delta\!\left(p_n-\frac
{q_{n-1}+q_n}2\right)\label{EWn}
\ee
We stress once more, however, that the regularized effective action \eqref{EW}
may not be 
derived only via \eqref{EWn}, that is via  the procedure of section 2.2. 
It could as well be obtained by means of the ordinary Feynman diagrams.

This amplitude has cyclic symmetry. When saturated with the
corresponding $h$'s,  as in \eqref{EW}, it gives the level $n$ effective action.
Here we would like to investigate some general consequences of the invariance of
the general effective action under the HS symmetry, codified by eq.
(\ref{deltahxp}), assuming for the $\EW^{(n)}$ the same cyclic symmetry as
\eqref{EWn}. The invariance of the effective action under \eqref{deltahxp} is
expressed as\footnote{Hereafter we assume that the HS symmetry is not anomalous
and that there is a regularization procedure leading to a HS invariant effective
action. The question of whether the particular effective action \eqref{Wreghm} 
satisfies (\ref{invariance}) requires an explicit calculation of \eqref{EWn} and
is left to future work.}
\be
0&=&\delta_\varepsilon W[h]\label{invariance}\\
&=&   \sum_{n=1}^\infty\, \frac 1{(n-1)!}\int
\prod_{i=1}^n d^dx_i\, \frac {d^dp_i}{(2\pi)^d}\, \0\\
&&\quad \times\, \EW_{\mu_1,\ldots, \mu_n}^{(n)}\!(x_1,p_1,\ldots, x_n, 
p_n,\epsilon)\, \delta_\varepsilon h^{\mu_1}(x_1,p_1) \ldots 
h^{\mu_n}(x_n,p_n)\0\\
&=&  \sum_{n=1}^\infty\, \frac 1{(n-1)!}\int \prod_{i=1}^n d^dx_i\, \frac
{d^dp_i}{(2\pi)^d}\, \0\\
&&\quad \times\,\EW^{(n)}_{\mu_1,\ldots, \mu_n}\!(x_1,p_1,\ldots, x_n, 
p_n)\,   
{\cal D}_x^{\ast\mu_1} \varepsilon(x_1,p_1)\, h^{\mu_2}(x_2,p_2) \ldots 
h^{\mu_n}(x_n,p_n)\0
\ee

In order to expose the $L_\infty$ structure we need the equations of motion
(EoM).
Here we can talk of {\it generalized equations of motion}. They are obtained by
varying
$W[h,\epsilon]$
with respect to $h^\mu(x,p)$:
\be
\frac {\delta }{\delta h^\mu(x,p)} W[h]=0\label{geneom}
\ee
Then, expanding in $p$, we obtain the generalized EoM's for the components
$h^{\mu_1\ldots\mu_n}(x)$. 

The most general EoM is therefore 
\be
\EF_\mu(x,p)=0\label{GenEoM}
\ee
where
\be \label{gEoMt}
\EF_\mu(x,p) &\equiv&  \sum_{n=0}^\infty\, \frac 1{n!}\int \prod_{i=1}^{n}
d^dx_i\,
\frac
{d^dp_i}{(2\pi)^d}\,  \EW_{\mu,\mu_1\ldots, \mu_{n}}^{(n+1)}(x,p,x_1,p_1,\ldots,
x_{n},  p_{n},\epsilon) \0\\
&&\quad \times \, h^{\mu_1}(x_1,p_1) \ldots  h^{\mu_{n}}(x_{n},p_{n})
\0
\ee 

Integrating by parts \eqref{invariance} { and using} (\ref{mypint}) 
we obtain the off-shell equation
\be
{\cal D}_x^{\ast\mu}\, \EF_\mu (x,p)\equiv \partial_x^\mu \EF_\mu(x,p) -i
[h^\mu (x,p)\stackrel{\ast}{,} \EF_\mu(x,p)]=0\label{DEF}
\ee
Taking the variation of this equation with respect to \eqref{deltahxp} we get
\be
0= \delta_\varepsilon ({\cal D}_x^{\ast\mu}\, \EF_\mu (x,p)) = {\cal
D}_x^{\ast\mu}\,\left( \delta_\varepsilon  \EF_\mu (x,p)\right)-i [  {\cal
D}_x^{\ast\mu}\varepsilon \stackrel{\ast}{,} \EF_\mu(x,p)]\label{dvareD}
\ee
From \eqref{DEF} and \eqref{dvareD} one can deduce
\be
\delta_\varepsilon \EF_\mu (x,p)=i [ \varepsilon(x,p) \stackrel{\ast}{,} 
\EF_\mu(x,p)]\label{dvareEF}
\ee

A final remark for this section. 
Using standard regularizations one obtains that in general the effective action 
contains term linear in HS fields, which gives constant contribution to EoM's
of 
even-spin HS fields of the form $c(s,\epsilon)\, (\eta_{\mu\mu})^{s/2}$, where
$c(s,\epsilon)$ are scheme dependent coefficients which need to be renormalized.
As this term is a generalization of the lowest-order contribution of the
cosmological constant term expanded around flat spacetime, we shall call the 
part of the effective action that contains the full linear term and is invariant
on HS transformations (\ref{deltahxp}), {\it generalized cosmological constant}
term. In the next section we shall assume that this term is removed from the
effective action.

\section{$L_\infty$ structure { in higher spin theory}}
\label{sec:Linfty}

\subsection{ $L_\infty$ symmetry of higher spin effective actions}

In this section we will uncover the $L_\infty$ symmetry of the $W[h]$. To this
end we use the general transformation properties derived in the previous
subsection, notably eqs. \eqref{GenEoM}, \eqref{dvareEF}, beside
\eqref{deltahxp}. 
We will also introduce a simplification, which is required by the classical form
of the  $L_\infty$ symmetry.
The expansion of the effective action (\ref{EW}) is in essence an expansion
around a flat background. As a flat background is not a solution when the
generalized cosmological constant term is present, consistency requires that we
take this term out of an effective action (or, in other words, renormalize the
cosmological constant to zero). This will be assumed from now on. Technically,
this means that we now assume that the sum in (\ref{EW})   starts from $n=2$,
and the sum in (\ref{gEoMt})   starts from $n=1$, while all other relations from
subsection \ref{ssec:gEoM}  are the same.

To start with let us recall that an $L_\infty$ structure characterizes closed
string field theory\footnote{Open
string field theory is instead characterized by an 
$A_\infty$ structure, see \cite{HZ} and references therein.}. This fact first
appeared in
\cite{Zwiebach}, see also \cite{Stasheff}, as a particular case of a general
mathematical structure 
called strongly homotopic algebras (or $SH$ algebras), see the introduction for 
physicists \cite{Lada1,Lada2}. It became later evident that this kind of
structure characterizes not only 
closed string field, but other field theories as well \cite{Barnich}, in
particular gauge field theories
\cite{Zeitlin}, Chern-Simons theories, Einstein gravity and double field theory
\cite{HZ}. For other more recent applications, see \cite{Gaiotto,Blumenhagen}. 

For the strongly homotopic algebra $L_\infty$ we closely follow the notation and
definitions of \cite{HZ}.
$L_\infty$ is determined by a set of vector spaces $X_i$, $i= \ldots,
1,0,-1,\ldots$, with degree $i$ and multilinear maps (products) among them 
$L_j$, $j=1,2,\ldots$, with degree $d_j= j-2$, satisfying the following
quadratic identities:
\be
\sum_{i+j=n+1} (-1)^{i(j-1)} \sum_\sigma (-1)^\sigma \epsilon(\sigma;x)\, L_j
(L_i(x_{\sigma(1)},\ldots, x_{\sigma(i)}),x_{\sigma(i+1)},\ldots
,x_{\sigma(n)})=0
\label{Ln}
\ee
In this formula $\sigma$ denotes a permutation of the entries so that $\sigma(1)<\ldots\sigma(i)$ and $\sigma(i+1)<\ldots\sigma(n)$, and
$\epsilon(\sigma;x)$ is the Koszul sign. To define it  consider an algebra with
product $x_i\wedge x_j=(-1)^{{\rm x}_i {\rm x}_j} {x_j\wedge x_i}$, where ${\rm
x}_i$ is the degree of $x_i$; then $\epsilon(\sigma;x)$ is defined by the
relation
\be
x_1\wedge x_2\wedge\ldots \wedge x_n = \epsilon(\sigma;x)\, x_{\sigma(1)}\wedge 
x_{\sigma(2)}\wedge \ldots\wedge x_{\sigma(n)}\label{koszul}
\ee

In our case, due to the structure of the effective action and the equation of
motion, we will need only three spaces 
$X_0,X_{-1},X_{-2}$ and the complex
\be
X_0 \stackrel{L_1}{\longrightarrow} X_{-1}
\stackrel{L_1}{\longrightarrow}X_{-2}\stackrel{L_1}{\longrightarrow}
0\label{Xcomplex}
\ee
The degree assignment is as follows: {$\varepsilon\in X_0$, $h^\mu\in X_{-1}$
and  ${\EF}_\mu \in X_{-2}$}.  

The properties of the mappings $L_i$ under permutation are defined in \cite{HZ}.
For instance
\be
L_2(x_1,x_2)&=& - (-1)^{\rm x_1 x_2}L_2(x_2,x_1)\label{genperm2}
\ee
In general 
\be 
 L_n(x_{\sigma(1)},x_{\sigma(2)},\ldots,x_{\sigma(n)})=(-1)^\sigma
\epsilon(\sigma;x) L_n(x_1,x_2,\ldots,x_n) \label{genpermn}
\ee
It is worth noting that if all the ${\rm x_i}$'s are odd 
$(-1)^\sigma \epsilon(\sigma;x)=1$.

The product $L_i$ are defined as follows. We first define the maps $\ell_i$ 
\be
\delta_\varepsilon h = \ell_1 (\varepsilon)+ \ell_2(\varepsilon,h)-\frac 12
\ell_3 (\varepsilon,h,h)-\frac 1{3!}  \ell_4
(\varepsilon,h,h,h)+\ldots\label{genlieh}
\ee
Therefore, in our case,
\be
\ell_1 (\varepsilon)^\mu&=& \partial_x^\mu \varepsilon(x,p)\label{ell1e}\\
\ell_2 (\varepsilon,h)^\mu&=& -i [h^\mu(x,p)\stackrel{\ast}{,}
\varepsilon(x,p)]=- \ell_2 (h,\varepsilon)^\mu\0\\
{\ell_j(\varepsilon,h,...,h)^\mu}&=& 0 \qquad,\qquad j \ge 3
\0
\ee
For these entries, i.e. $\varepsilon,(\varepsilon,h), (\varepsilon,h,h),\ldots$
we set $L_i=\ell_i$.

From the above we can extract
$L_2(\varepsilon,\varepsilon)\equiv\ell_2(\varepsilon,\varepsilon)$. We have
\be
\left(\delta_{\varepsilon_1}
\delta_{\varepsilon_2}-\delta_{\varepsilon_2}\delta_{\varepsilon_1}
\right)h^\mu&=&
\delta_{\varepsilon_1}\left( \ell_1(\varepsilon_2)+ \ell_2
(\varepsilon_2,h)\right)-
\delta_{\varepsilon_2}\left( \ell_1(\varepsilon_1)+ \ell_2
(\varepsilon_1,h)\right)\label{deltae1e2'}\\
&=&\delta_{\varepsilon_1}\left(   \ell_2 (\varepsilon_2,h)\right)-
\delta_{\varepsilon_2}\left(   \ell_2 (\varepsilon_1,h)\right)\0\\
&=&  \ell_2 (\varepsilon_2,\delta_{\varepsilon_1}h)-  \ell_2
(\varepsilon_1,\delta_{\varepsilon_2}h) =  \ell_2
(\varepsilon_2,\ell_1(\varepsilon_1))- \ell_2
(\varepsilon_1,\ell_1(\varepsilon_2))+ {\cal O}(h)\0
\ee

Now, the $L_\infty$ relation (\ref{Ln}) involving $L_1$ and $L_2$ is 
\be
L_1 (L_2(x_1,x_2) )= L_2(L_1(x_1),x_2)- (-1)^{{\rm x_1}{\rm x_2}} 
L_2(L_1(x_2),x_1)
\label{l1l2}
\ee
for two generic elements of $x_1,x_2$ of degree ${\rm x_1},{\rm x_2}$,
respectively. If we wish to satisfy it we have to identify
\be
\left(\delta_{\varepsilon_1}
\delta_{\varepsilon_2}-\delta_{\varepsilon_2}\delta_{\varepsilon_1}
\right)h&=&-\ell_1(\ell_2(\varepsilon_1,\varepsilon_2))+ {\cal
O}(h)\label{deltae1e2''}
\ee
By comparing this with (\ref{e1e2}) we obtain
\be
{ \ell_2(\varepsilon_1,\varepsilon_2)=  i\, [{\varepsilon_1}\stackrel{\ast}{,}
{\varepsilon_2}]}
\label{l2e1e2}
\ee

The next step is to determine $L_3$. It must satisfy, in particular, the
$L_\infty$ relation
\be
0&=& L_1(L_3(x_1,x_2,x_3))\label{L1L3}\\
&+& L_3(L_1(x_1),x_2,x_3)+ (-1)^{\rm x_1}L_3(x_1,L_1(x_2),x_3)+
(-1)^{\rm x_1+x_2}L_3(x_1,x_2,L_1(x_3))\0\\
&+&L_2(L_2(x_1,x_2),x_3)+(-1)^{\rm
(x_1+x_2)x_3}L_2(L_2(x_3,x_1),x_2)+(-1)^{\rm
(x_2+x_3)x_1}L_2(L_2(x_2,x_3),x_1)\0
\ee

We define first the $\ell_i$ with only $h$ entries. They are given by the {generalized} EoM:
\be
\EF = \ell_1(h) -\frac 12  \ell_2(h,h) -\frac 1{3!}
\ell_3(h,h,h)+\ldots\label{genlihh}
\ee 
Let us write $\EF_\mu$, \eqref{GenEoM} in compact form as
\be
\EF_\mu = \sum_{n=1}^\infty  \frac 1{n!} \langle\!\langle \EW_\mu^{(n+1)} ,
h^{\otimes n}\rangle\!\rangle 
\label{GenEoMcompact}
\ee
then 
\be
\ell_n(h,\ldots,h)&=&(-1)^{\frac{n(n-1)}2} \langle\!\langle \EW_\mu^{(n+1)} ,
h^{\otimes n}\rangle\!\rangle\label{ellnhhh}\\
&=& (-1)^{\frac{n(n-1)}2} \int \prod_{i=1}^{n} d^dx_i\, \frac
{d^dp_i}{(2\pi)^d}\,  \EW_{\mu,\mu_1\ldots, \mu_{n}}^{(n+1)}(x,p,x_1,p_1,\ldots,
x_{n},  p_{n})\0\\
&&\times\,  h^{\mu_1}(x_1,p_1) \ldots  h^{\mu_{n}}(x_{n},p_{n})\0
\ee
{ in particular, \be
\ell_1(h) = \langle\!\langle \EW_\mu^{(2)} ,h\rangle\!\rangle = 
\int  d^dx_i\, \frac
{d^dp_i}{(2\pi)^d}\,  \EW_{\mu,\mu_1}^{(2)}(x,p,x_1,p_1)
 h^{\mu_1}(x_1,p_1)\label{ell1h}
\ee}
Notice that $ \EW_{\mu,\mu_1\ldots, \mu_{n}}^{(n+1)}$ is not symmetric in the
exchange of its indices. 
In fact it has only a cyclic symmetry.  But in order to verify
the $L_\infty$ relations 
we have to know these products for different entries. Following \cite{HZ} we
define, for instance,
\be
2L_2(h_1,h_2)&=& \ell_2(h_1+h_2,h_1+h_2) -
\ell_2(h_1,h_1)-\ell_2(h_2,h_2)\label{lh1h2}\\
\ee
which is equivalent to
\be
L_2(h_1,h_2)&=&\frac 12 \left( \ell_2(h_1,h_2) +
\ell_2(h_2,h_1)\right)
\label{L2}
\ee
Similarly
\be
L_3(h_1,h_2,h_3)= \frac 16 \left(
\ell_3(h_1,h_2,h_3)+ {\rm perm}(h_1,h_2,h_3)\right)\label{L3}
\ee
In  general, when we have a non-symmetric $n$-linear function $f_n$ of the
variable $h$ we can generate a 
symmetric function $F_n$ linearly dependent on each of $n$ variables
$h_1,\ldots, h_n$ through the following process
\be 
&&F_n(h_1,\ldots, h_n)\0\\
&&= \frac 1{n!} \Big{(} f_n(h_1+ \ldots +h_n) -\Big{[}f_n(h_1+\ldots +h_{n-1})
+ f_n(h_1+\ldots +h_{n-2}+h_n)\0\\
&&\quad+ \ldots +f_n(h_2+\ldots +h_n) \Big{]} 
+ \Big{[} f_n(h_1+\ldots+ h_{n-2})+ \dots+f_n(h_3+\ldots+h_n)\Big{]}\0\\
&&\quad+\ldots\0\\
&&\quad+ (-1)^{n-k} \Big{[} f_n(h_1+\ldots+ h_{k})+
\dots+f_n(h_{n-k+1}+\ldots+h_n)\Big{]}\0\\
&&\quad+\ldots\0\\
&&\quad+ (-1)^{n-1} \Big{[} f_n(h_1) +\ldots + f_n(h_n)\Big{]}
\label{symmbydiag}
\ee
{ We shall define $L_n(h_1,\ldots,h_n)$ by using this formula: replace $F_n$ with
$L_n$ and $f_n$ with $\ell_n$,
the latter being given by (\ref{ellnhhh}).}

{ We shall see that beside $L_n(h_1,\ldots,h_n)$, (\ref{ell1e}) and (\ref{l2e1e2}) 
the only nonvanishing objects defining the $L_\infty$ algebra of the HS}
effective action are
\be
L_2(\varepsilon,E) = i [\varepsilon \stackrel{\ast}{,} E]
\ee
where $E$ represents $\mathcal{F}_\mu$ or any of its homogeneous  {pieces}.

In the rest of this section we shall prove that $L_n$ defined in this way
generate an $L_\infty$ 
algebra.

\subsection{Proof of the $L_\infty$ relations}

\subsubsection{Relation $L_1^2=0$, degree -2}

Now let us verify the remaining $L_\infty$ relations. The first is $L_1^2\equiv
\ell_1^2=0$. \footnote{We remark that if the generalized cosmological constant
term (see end of sec.\ \ref{ssec:gEoM}   and beginning of sec.\
\ref{sec:Linfty})  is non-vanishing, then $\ell_1^2\neq 0$. In this case an
enlarged version of $L_\infty$, {called {\it curved} $L_\infty$}, is necessary. We thank J. Stasheff for this piece of information. We will not explore this possibility here.}
 
Let us start from $\ell_1 (\ell_1(\varepsilon)) $. We recall that
$\ell_1(\varepsilon)=
\partial_x \varepsilon(x,p)$ and belongs to $X_{-1}$. Now
\be 
\ell_1(h)= \langle\!\langle \EW_\mu^{(2)} , h\rangle\!\rangle\label{ell1h}
\ee
Replacing $h$ with $\partial_x \varepsilon(x,p)$ corresponds to taking the
variation of the lowest order in $h$ of $\EF_\mu$ with respect to $h$, i.e. with
respect to \eqref{deltahxp}. On the other hand the variation of $\EF_\mu$ is
given by \eqref{dvareEF} and is linear in  $\EF_\mu$. Therefore, since
$\ell_1(\partial_x \varepsilon(x,p))$ is order 0 in $h$ it must vanish. In fact
it does, which corresponds to the gauge invariance of the EoM to the lowest
order in $h$.

Next let us consider $\ell_1(\ell_1(h))$. It has degree -3, so it is {necessarily 0 since
$X_{-3}=0$.}

\subsubsection{Relation $L_1L_2=L_2L_1$, degree -1}

Next, we know $\ell_2(\varepsilon_1,\varepsilon_2), \ell_2 (\varepsilon, h)$ and
$ \ell_2(h_1,h_2)$, and we have to verify
$L_1L_2=L_2L_1$. The latter { is written explicitly in} (\ref{l1l2}) and takes
the form
\be
\ell_1(\ell_2(\varepsilon,h))&=&
L_2(\ell_1(\varepsilon),h) {+}
L_2(\varepsilon,\ell_1(h))\label{ell1ell2}\\
&=&\frac 12 \Big{(}
\ell_2(\ell_1(\varepsilon),h)+\ell_2(h,\ell_1(\varepsilon))\Big{)}
 {+} L_2 (\varepsilon, \ell_1(h))\0
\ee
{ where we used} (\ref{L2}).
More explicitly \eqref{ell1ell2} writes
\be 
-i \ell_1([h\stackrel{\ast}{,} \varepsilon])_\mu=\frac 12 \Big{(}
\ell_2(\partial^x\varepsilon, h)
+\ell_2(h,\partial^x\varepsilon)\Big{)}_\mu + L_2 (\varepsilon, \langle\!\langle
\EW_\mu^{(2)},h\rangle\!\rangle)\label{ell1ell2b}
\ee
i.e.
\be 
i \langle\!\langle \EW_{\mu\nu}^{(2)}\, ,\,[h^\nu\stackrel{\ast}{,}
\varepsilon])\rangle\!\rangle = \frac 12 \Big{(} \langle\!\langle
\EW_{\mu\nu\lambda}^{(3)}\, , \,
\partial_x^\nu \varepsilon \, h^\lambda\rangle\!\rangle+ \langle\!\langle
\EW_{\mu\nu\lambda}^{(3)}\, , \,
h^\nu \,\partial_x^\lambda \varepsilon\rangle\!\rangle\Big{)}
- L_2 (\varepsilon, \langle\!\langle \EW_\mu^{(2)},h\rangle\!\rangle) \qquad
\label{ell1ell2c}
\ee 
To understand this relation one must unfold \eqref{dvareEF}. On one side we have
\be 
\delta_\varepsilon \EF_\mu&=& \sum_{n=1}^\infty
\frac 1{n!}\Big{(} \sum_{i=1}^{n} \langle\!\langle \EW_{\mu\mu_1\ldots
\mu_i\ldots \mu_{n}}^{(n+1)}\,
,\,h^{\mu_1}\ldots\partial_x^{\mu_i}\varepsilon\,  \ldots h^{\mu_{n} }
\rangle\!\rangle \label{deltaEF1}\\
&&-i \, \sum_{i=1}^{n} 
\langle\!\langle \EW_{\mu\mu_1\ldots \mu_i\ldots \mu_n}^{(n+1)}\, ,\,
h^{\mu_1}\ldots\,[h^{\mu_i}\stackrel{\ast}{,}
\varepsilon]\ldots h^{\mu_n}\rangle\!\rangle\Big{)} \0
\ee
On the other side
\be
i[\varepsilon\stackrel{\ast}{,} \EF_\mu] = i \sum_{n=1}^\infty \frac 1{n!}
[\varepsilon \stackrel{\ast}{,}\langle\!\langle \EW_{\mu}^{(n+1)}\, ,\,
h^{\otimes n}\rangle\!\rangle]\label{deltaEF2}
\ee
The two must be equal order by order in $h$. Thus we have
\be
i [\varepsilon \stackrel{\ast}{,}\langle\!\langle \EW_{\mu}^{(n+1)}\, ,\,
h^{\otimes n}\rangle\!\rangle]&=&\frac 1{n+1}\sum_{i=1}^{n+1} \langle\!\langle
\EW_{\mu\mu_1\ldots
\mu_i\ldots \mu_{n+1}}^{(n+2)}\,
,\,h^{\mu_1}\ldots\partial_x^{\mu_i}\varepsilon\,  \ldots h^{\mu_{n+1} }
\rangle\!\rangle\label{EFequations}\\
&& - i \, \sum_{i=1}^{n} 
\langle\!\langle \EW_{\mu\mu_1\ldots \mu_i\ldots \mu_n}^{(n+1)}\, ,\,
h^{\mu_1}\ldots\,[h^{\mu_i}\stackrel{\ast}{,}
\varepsilon]\ldots h^{\mu_n}\rangle\!\rangle \0
\ee
This is a not too disguised form of the Ward identity for the symmetry
\eqref{deltahxp}.
Setting $n=1$ gives precisely \eqref{ell1ell2c} provided
\be
L_2 (\varepsilon, \langle\!\langle
\EW_\mu^{(2)},h\rangle\!\rangle)=i  [\varepsilon
\stackrel{\ast}{,} \langle\!\langle\EW_\mu^{(2)}\, ,\,
h\rangle\!\rangle]\label{EW2}
\ee
The quantity $\mathcal{F}^{(1)}=  \langle\!\langle\EW_\mu^{(2)}\,
,\,h\rangle\!\rangle$ is
the lowest order piece of the EoM (of degree -2), see (\ref{GenEoMcompact}).  
So we can say
\be
L_2 (\varepsilon,\mathcal{F}^{(1)})\equiv \ell_2
(\varepsilon,\mathcal{F}^{(1)})= i[\varepsilon
\stackrel{\ast}{,} \mathcal{F}^{(1)}] \label{ell2E1}
\ee
{In general,
\be
\ell_2
(\varepsilon,\mathcal{F})= i[\varepsilon
\stackrel{\ast}{,} \mathcal{F}] \label{ell2E}
\ee}

The next relation to be verified is 
\be
L_1(L_2(h_1,h_2))= L_2(L_1(h_1),h_2)-
L_2(h_1,L_1(h_2))\label{ell1ell2hh}
\ee
The entries of $L_2$ on the rhs have degree -3, so they must vanish. On the
other hand $L_2(h_1,h_2)$ on the lhs has degree -2, and is mapped to degree
-3 by $L_1$. So it is consistent to equate both sides to 0. In particular we
can set $L_2(\mathcal{F}^{(1)},h)=0$ {(and, more generally, $L_2(X_{-2},h)=0$).}

\subsubsection{Relation $L_3L_1 +L_2L_2+L_1L_3=0 $, degree 0}

First we should evaluate $L_3(\varepsilon_1,\varepsilon_2,\varepsilon_3)$. Its
degree is 1, therefore
it exits the complex. Is it consistent to set it to 0? The relevant $L_\infty$
relation is
\be
0&=& \ell_1(L_3(x_1,x_2,x_3))\label{ell1ell3}\\
&+& L_3(\ell_1(x_1),x_2,x_3)+ (-1)^{\rm x_1}L_3(x_1,\ell_1(x_2),x_3)+
(-1)^{\rm x_1+x_2}L_3(x_1,x_2,\ell_1(x_3))\0\\
&+&{L}_2({L}_2(x_1,x_2),x_3)+(-1)^{\rm
(x_1+x_2)x_3}{L}_2({L}_2(x_3,x_1),x_2)+(-1)^{\rm
(x_2+x_3)x_1}{L}_2({L}_2(x_2,x_3),x_1)\0
\ee
In our case the second line equals $\partial_x
L_3(\varepsilon_1,\varepsilon_2,\varepsilon_3)$. Thus if we set 
$L_3(\varepsilon_1,\varepsilon_2,\varepsilon_3)=0$, the first  two lines
vanish.  Using \eqref{l2e1e2}, we see that the third line is nothing but the
$\ast$-Jacobi identity.

Arguing the same way and using the next $L_\infty$ relation, which involves
$L_4$, one can show that
  $L_4(\varepsilon_1,\varepsilon_2,\varepsilon_3,\varepsilon_4)=0$, etc.

From \eqref{ell1e} we also know that 
$L_3(\varepsilon,h_1,h_2)\equiv\ell_3(\varepsilon,h_1,h_2)=0$.
Following \cite{HZ} we will set also
$L_3(\varepsilon_1,\varepsilon_2,h)=0$,  
$L_3(\varepsilon_1,\varepsilon_2,\mathcal{F}^{(1)})=0$. Therefore
\be
L_3(\varepsilon_1,\varepsilon_2,\varepsilon_3)=0,\quad
L_3(\varepsilon,h_1,h_2)=0,\quad
L_3(\varepsilon_1,\varepsilon_2,h)=0,
\quad L_3(\varepsilon_1,\varepsilon_2,\mathcal{F}^{(1)})=0\label{ident0}
\ee

Let us consider next the entries
$\varepsilon_1,\varepsilon_2,h$. The terms of the first two lines in
\eqref{L1L3} vanish due to \eqref{ident0}. The last line is
\be
&&\ell_2(\ell_2( \varepsilon_1,\varepsilon_2),h) +\ell_2(\ell_2(h,
\varepsilon_1),\varepsilon_2)+ \ell_2(\ell_2(
\varepsilon_2,h),\varepsilon_1)\0\\
&&= [h^\mu \stackrel{\ast}{,} [\varepsilon_1\stackrel{\ast}{,}\varepsilon_2]] 
-[[h^\mu \stackrel{\ast}{,}
\varepsilon_1]\stackrel{\ast}{,}\varepsilon_2]+[[h^\mu \stackrel{\ast}{,}
\varepsilon_2]\stackrel{\ast}{,}\varepsilon_1]
\ee
which vanishes due to $\ast$-Jacobi identity.

Now we consider the entries $\varepsilon ,h_1,h_2$. Plugging them into
\eqref{L1L3},  the first line vanishes because of \eqref{ident0}. The rest is
\be
0&=& \frac 16 \Big{(}{\ell_3(\ell_1(\varepsilon),h_1,h_2)+{\rm
perm}_3}\Big{)}\0\\
&&+ L_3(\varepsilon,\ell_1(h_1),h_2) -L_3(\varepsilon,h_1,\ell_1(h_2)) \0\\
&&+\frac 12 \Big{(} \ell_2 (\ell_2(\varepsilon,h_1),h_2)+ \ell_2
(h_2,\ell_2(\varepsilon,h_1)) -\ell_2 (\ell_2(h_2,\varepsilon),h_1)\0\\
&&~~~~~~- \ell_2 (h_1,\ell(h_2,\varepsilon)) +
\ell_2 (\ell_2( h_1,h_2),\varepsilon)+\ell_2(\ell_2(
h_2,h_1),\varepsilon)\Big{)}\label{ellepsh1h2}
\ee
where ${\rm perm}_3$ means the permutation of the three entries of $\ell_3$.
Writing down explicitly the first line, it takes the form
\be
\frac 16 \left(\ell_3(\ell_1(\varepsilon),h_1,h_2) +{\rm perm}_3\right) =-\frac
16 \left( \langle\!\langle
\EW_{\mu\nu\lambda\rho}^{(4)}\, ,\, \partial_x^\nu \varepsilon \, h_1^\lambda
\,h_2^\rho\rangle\!\rangle +{\rm perm}_3\right)   \label{ell3h1h2first}
\ee
The last two lines of \eqref{ellepsh1h2} give
\be
&&\ell_2 (\ell_2(\varepsilon,h_1),h_2)+ \ell_2 (h_2,\ell_2(\varepsilon,h_1))
-\ell_2 (\ell_2(h_2,\varepsilon),h_1)- \ell_2 (h_1,\ell_2(h_2,\varepsilon)) +
\ell_2 (\ell_2( h_1,h_2),\varepsilon) \0\\
&&+{\ell_2(\ell_2( h_2,h_1),\varepsilon)} = {+}i\Big{(} 
\langle\!\langle\EW_{\mu\nu\lambda}^{(3)}\, ,\,
[h_1^\nu
\stackrel{\ast}{,} \varepsilon] h_2^\lambda \rangle\!\rangle
+\langle\!\langle\EW_{\mu\nu\lambda}^{(3)}\, ,\,h_2^\lambda  [h_1^\nu
\stackrel{\ast}{,} \varepsilon] \rangle\!\rangle 
+ \langle\!\langle\EW_{\mu\nu\lambda}^{(3)}\, ,\, [h_2^\nu \stackrel{\ast}{,}
\varepsilon] h_1^\lambda \rangle\!\rangle \0\\
&& + \langle\!\langle\EW_{\mu\nu\lambda}^{(3)}\, ,\, h_1^\lambda  [h_2^\nu
\stackrel{\ast}{,}
\varepsilon]\rangle\!\rangle {+
[ \varepsilon \stackrel{\ast}{,} \langle\!\langle\EW_{\mu\nu\lambda}^{(3)}\, ,\, h_1^\nu\, h_2^\lambda
\rangle\!\rangle ]+[\varepsilon
 \stackrel{\ast}{,}\langle\!\langle\EW_{\mu\nu\lambda}^{(3)}\, ,\,h_2^\lambda\, h_1^\nu
\rangle\!\rangle  ] }\Big{)}\label{ell3h1h2second}
\ee
Summing the rhs's of \eqref{ell3h1h2first} and \eqref{ell3h1h2second} one gets, 
apart from the second line, \eqref{ellepsh1h2} expressed in terms of  the
expressions appearing 
in the rhs of \eqref{EFequations} with entries $h_1,h_2 $, instead of one single
$h$.

Now let us consider \eqref{EFequations} for $n=2$, i.e. 
\be
i [\varepsilon \stackrel{\ast}{,}  \langle\!\langle\EW_{\mu\nu\lambda}^{(3)}\,
,\, h^\nu\, h^\lambda \rangle\!\rangle] &=&\frac 13
\langle\!\langle \EW_{\mu\nu\lambda\rho}^{(4)}\, ,\,\partial_x^\nu\varepsilon
h^\lambda h^\rho+ h^\nu \partial_x^\lambda \varepsilon h^\rho+ 
\, h^\nu\,h^\lambda \partial^\rho_x \varepsilon
\rangle\!\rangle\label{EFequationsn=2}\\
&& -i\, \langle\!\langle\EW_{\mu\nu\lambda}^{(3)}\, ,\, [h^\nu
\stackrel{\ast}{,} \varepsilon] h^\lambda + h^\nu [h^\lambda
\stackrel{\ast}{,} \varepsilon] \rangle\!\rangle\0.
\ee
This can be read as
\be
-i [\varepsilon \stackrel{\ast}{,} \ell_2(h,h) ] &=& -\frac 13
\Big{(}\ell_3(\partial_x\varepsilon, h,h)+ \ell_3(h,\partial_x\varepsilon,
h)+\ell_3(h,h,\partial_x\varepsilon)\Big{)}\0\\
&&+ i \ell_2(h,[h\stackrel{\ast}{,} \varepsilon])+i \ell_2([h\stackrel{\ast}{,}
\varepsilon],h)\label{varepsilonhh}
\ee
Now we consider the same equation obtained by replacing $h$ with $h_1+h_2$
according to the symmetrization procedure in \eqref{lh1h2}. We get in this way
the symmetrized equation
\be
&& -i [\varepsilon \stackrel{\ast}{,} \ell_2(h_1,h_2) ] - i [\varepsilon
\stackrel{\ast}{,} \ell_2(h_2,h_1) ]  \0\\
&=& -\frac 13 \Big{(}\ell_3(\partial_x\varepsilon, h_1,h_2)+
\ell_3(\partial_x\varepsilon, h_2,h_1)+\ell_3(h_1,\partial_x\varepsilon,
h_2)\0\\
&&+\ell_3(h_2,\partial_x\varepsilon,
h_1)+\ell_3(h_1,h_2,\partial_x\varepsilon)+\ell_3(h_2,h_1,
\partial_x\varepsilon)\Big{)}\0\\
&&+ i \ell_2(h_1,[h_2\stackrel{\ast}{,} \varepsilon])+ i
\ell_2(h_2,[h_1\stackrel{\ast}{,} \varepsilon])+i \ell_2([h_1\stackrel{\ast}{,}
\varepsilon],h_2)+i \ell_2([h_2\stackrel{\ast}{,}
\varepsilon],h_1)\label{varepsilonhh1}
\ee
This is the same as the sum of the first, third and fourth lines of 
\eqref{ellepsh1h2}, or, alternatively, the sum of the  rhs's of
\eqref{ell3h1h2first} and \eqref{ell3h1h2second}.

Thus \eqref{ellepsh1h2} is satisfied if the two remaining terms in the second
line vanish. They are all of the type $L_3(\varepsilon, h,\mathcal{F}^{(1)})$
and we can
assume that such types of terms vanish. So, beside \eqref{ident0} we have
\be
 L_3(\varepsilon, h,E)=- L_3(\varepsilon, E,h) =0\label{ident0a}
\ee
{ where $E$ represent $\EF_\mu$ {or anything in $X_{-2}$.}

The relation with entries  $\varepsilon_1,\varepsilon_2$ 
and $E$ is nontrivial and has to be verified. Consider
again \eqref{L1L3} with entries $\varepsilon_1,\varepsilon_2$ 
and $E$. Due to \eqref{ident0},\eqref{ident0a} the relation \eqref{L1L3} 
reduces to the last line:
\be 
&&\ell_2(\ell_2(\varepsilon_1,\varepsilon_2),E)+
\ell_2(\ell_2(E,\varepsilon_1),\varepsilon_2)+ 
\ell_2(\ell_2(\varepsilon_2,E),\varepsilon_1)\label{ell2ell2E}\\
&& =i \ell_2( [\varepsilon_1\stackrel{\ast}{,}   \varepsilon_2],E) +i
\ell_2([E\stackrel{\ast}{,} \varepsilon_1],
\varepsilon_2)+i\ell_2([\varepsilon_2\stackrel{\ast}{,} E], \varepsilon_1)\0\\
&&= {+}[E\stackrel{\ast}{,} [\varepsilon_1\stackrel{\ast}{,} \varepsilon_2] 
{]-} [[E\stackrel{\ast}{,} \varepsilon_1]\stackrel{\ast}{,}  \varepsilon_2]{-}
[[\varepsilon_2\stackrel{\ast}{,} E]\stackrel{\ast}{,} {\varepsilon_1}]\0
\ee
which vanishes because of the $\ast$-Jacobi identity.

\subsubsection{Relation $L_1L_4 -L_2L_3+L_3L_2-L_4L_1 =0 $, degree 1}

The $L_\infty$ relation to be proved at degree 1 is
\be
&& L_1(L_4(x_1,x_2,x_3,x_4))\label{L1L4}\\
&&-L_2( L_3(x_1,x_2,x_3),x_4)+(-1)^{\rm x_3x_4}L_2( L_3(x_1,x_2,x_4),x_3)\0\\
&&+(-1)^{(1+\rm x_1)\rm x_2}L_2(x_2, L_3(x_1,x_3,x_4))-(-1)^{\rm x_1} L_2(x_1,
L_3(x_2,x_3,x_4))\0\\
&&+ L_3(L_2(x_1,x_2),x_3,x_4)+ (-1)^{1+\rm x_2 x_3}L_3(L_2(x_1,x_3),x_2,x_4)\0\\
&&+
(-1)^{\rm x_4(x_2+x_3)}L_3(L_2(x_1,x_4),x_2,x_3)\0\\
&&-L_3(x_1,L_2(x_2,x_3),x_4)+(-1)^{\rm
x_3x_4}L_3(x_1,L_2(x_2,x_4),x_3)+L_3(x_1,x_2,L_2(x_3,x_4))\0\\
&&- L_4(L_1(x_1),x_2,x_3,x_4)- (-1)^{\rm x_1}L_4(x_1,L_1(x_2),x_3,x_4)\0\\
&&-(-1)^{\rm x_1+x_2}L_4(x_1,x_2,L_1(x_3),x_4)- (-1)^{\rm x_1+x_2+x_4}
L_4(x_1,x_2,x_3,L_1(x_4))=0\0
\ee
We have
\be
L_4(\varepsilon_1,\varepsilon_2,\varepsilon_3,\varepsilon_4)=0,\quad
L_4(\varepsilon_1,\varepsilon_2,\varepsilon_3,h)=0,\quad
L_4(\varepsilon_1,\varepsilon_2,h_1,h_2)=0,\quad
L_4(\varepsilon,h_1,h_2,h_3)=0\0\\
\label{ident04}
\ee
The first and second equality have positive degree, so they must vanish. The
fourth has
been proven above, see \eqref{ell1e}. The other is an ansatz to be checked by
consistency.

The relation \eqref{L1L4} with four $\varepsilon$ entries has already been
commented. The same relation with three $\varepsilon$ entries and
one $h$ is also trivial as a consequence of \eqref{ident0} and \eqref{ident04}.
The same happens in the case of two $\varepsilon$ entries and
two $h$, as a consequence again of \eqref{ident0} and \eqref{ident04}. 

Now let us consider the case of one $\varepsilon$ and three $h$'s. Plugging them
into \eqref{L1L4} here is what we get in terms of $\ell_i$'s
(only the nonzero terms are written down)
\be
0&=& -\frac 16 \Big{(}\ell_2 (\varepsilon, \ell_3(h_1,h_2,h_3)) + {\rm perm}_3 
\Big{)}\label{L4a}\\
&&+\frac 16 \Big{(} \ell_3(\ell_2(\varepsilon, h_1),h_2,h_3) +
 \ell_3(\ell_2(\varepsilon,
h_2),h_1,h_3)   +  \ell_3(\ell_2(\varepsilon,
h_3),h_1,h_2) +{\rm perm}_3\Big{)}\0\\
&&-\frac 1{4!} \Big{(}\ell_4(\ell_1(\varepsilon), h_1,h_2,h_3)+ {\rm perm}_4
 \Big{)}\0\\
&&{-} L_4(\varepsilon, \ell_1(h_1),h_2,h_3)+L_4(\varepsilon,h_1, \ell_1(h_2),h_3)
{-}L_4(\varepsilon,h_1, h_2,\ell_1(h_3))\0
\ee  
where ${\rm perm}_3, {\rm perm}_4$ refer to the permutations of the $\ell_3,
\ell_4$ entries, respectively.
Disregarding for the moment the last line, which is of type
$L_4(\varepsilon,E,h,h)$, this equation
becomes
\be
0&=& \frac i6 \Big{(}[\varepsilon \stackrel{\ast}{,} \langle\!\langle
\EW^{(4)}_{\mu\nu\lambda\rho}\, ,\,h_1^\nu h_2^\lambda h_3^\rho
\rangle\!\rangle]+
{\rm perm}(h_1,h_2,h_3)\label{L4b}\\
&& +\langle\!\langle \EW^{(4)}_{\mu\nu\lambda\rho}\, ,\, [h_1^\nu
\stackrel{\ast}{,} \varepsilon ] h_2^\lambda h_3^\rho\rangle\!\rangle +{\rm
perm}( [h_1
\stackrel{\ast}{,} \varepsilon ] ,h_2 ,h_3)\0\\
&&+  \langle\!\langle \EW^{(4)}_{\mu\nu\lambda\rho}\, ,\, [h_2^\nu
\stackrel{\ast}{,} \varepsilon ] h_1^\lambda h_3^\rho\rangle\!\rangle +{\rm
perm}([h_2
\stackrel{\ast}{,} \varepsilon ], h_1,h_3)\0\\
&& + \langle\!\langle \EW^{(4)}_{\mu\nu\lambda\rho}\, ,\, [h_3^\nu
\stackrel{\ast}{,} \varepsilon ] h_1^\lambda h_2^\rho\rangle\!\rangle +{\rm
perm}([h_3
\stackrel{\ast}{,} \varepsilon ], h_1, h_2)\Big{)} \0\\
&&-\frac 1{4!}\Big{(}\langle\!\langle \EW^{(5)}_{\mu\nu\lambda\rho\sigma}\, ,\, 
\partial_x^\nu\varepsilon\,  h_1^\lambda h_2^\rho h_3^\sigma\rangle\!\rangle  +
{\rm perm}
(\partial_x\varepsilon,h_1,h_2,h_3)\Big{)}\0
\ee 
For comparison let us go back to \eqref{EFequations} with $n=3$. It writes
\be
&&i[\varepsilon \stackrel{\ast}{,} \langle\!\langle
\EW^{(4)}_{\mu\nu\lambda\rho}\, ,\,h^\nu h^\lambda h^\rho \rangle\!\rangle]\0\\
&&=\frac 14
\langle\!\langle \EW^{(5)}_{\mu\nu\lambda\rho\sigma}\, ,\,
\partial_x^\nu\varepsilon 
h^\lambda h^\rho h^\sigma+ h^\nu \partial_x^\lambda\varepsilon 
h^\rho h^\sigma+ h^\nu h^\lambda  \partial_x^\rho\varepsilon h^\sigma+ h^\nu
h^\lambda h^\rho  \partial_x^\sigma\varepsilon\rangle\!\rangle \0\\
&&-\, i
\langle\!\langle \EW^{(4)}_{\mu\nu\lambda\rho}\, ,\, [h^\nu \stackrel{\ast}{,}
\varepsilon ] h^\lambda h^\rho+ h^\nu [h^\lambda\stackrel{\ast}{,}
\varepsilon ] h^\rho  + h^\nu  h^\lambda  [h^\rho\stackrel{\ast}{,}
\varepsilon ]\rangle\!\rangle \label{EFequationsn=3}
\ee
If now we transform the LHS of this equation to a trilinear function of
$h_1,h_2, h_3$ according to the recipe \eqref{symmbydiag}, we obtain precisely 
eq.\eqref{L4b}. As a consequence we are forced to set
\be
L_4(\varepsilon, E,h,h)=L_4(\varepsilon, h,E,h)=L_4(\varepsilon,
h,h,E)=0\label{ident04b}
\ee 
Considering the entries $\varepsilon,\varepsilon,E,h$ in \eqref{L1L4} one can
show that
\be
L_4(\varepsilon,\varepsilon,E,h)=0\label{indet04c}
\ee
for consistency. Using this and evaluating \eqref{L1L4} with entries 
$\varepsilon,\varepsilon,h,h$, one can see that the third ansatz in 
\eqref{ident04} is justified.

\subsubsection{Relation $L_1L_n+\ldots \pm L_n L_1 =0 $, degree $n-3$}

The general $L_\infty$ relation is \eqref{Ln}.
 As the $n=4$ example shows, for $n\geq 4$ it is consistent to set the values of
$L_n$ 
to zero except when all the entries have degree -1. Schematically, out of
\eqref{Ln}, the only nontrivial relation is
\be
-L_2(\varepsilon, L_{n-1}(h,\ldots,h)) + L_{n-1}(L_2(\varepsilon,h),h,\ldots,h)
+(-1)^{n-1}L_n(L_1(\varepsilon),h,\ldots,h)=0\0\\
\label{Lnrel}
\ee
Written in explicit form in terms of $\ell_n$, it is
\be
&&-\frac {1}{(n-1)!}\Big{(}\ell_2(\varepsilon,\ell_{n-1}(h_1,\ldots,h_{n-1})) +
{\rm perm}_{n-1} \Big{)}\label{Lnrel1}\\
&&+\frac{1}{(n-1)!}
\Big{(}\ell_{n-1}(\ell_2(\varepsilon,h_1),h_2,\ldots,h_{n-1}) 
+\ell_{n-1}(\ell_2(\varepsilon,h_2),h_1,\ldots,h_{n-1})+\ldots \0\\
&&\quad\quad\quad\quad\quad+
\ell_{n-1}(\ell_2(\varepsilon,h_{n-1}),h_1,\ldots,h_{n-2}) +{\rm perm}_{n-1}
\Big{)} \0\\ 
&& + \frac {(-1)^{n-1}}{n!}\Big{(}\ell_n(\ell_1(\varepsilon), h_1,\ldots,
h_{n-1})+{\rm perm}_{n} \Big{)} =0\0 
\ee
In order to obtain this it is essential to remark that, for entries of degree
-1, the factor $ (-1)^\sigma \epsilon(\sigma;x)$ in \eqref{Ln} is 1.

Using now the definition \eqref{ellnhhh} and simplifying, \eqref{Lnrel1} becomes
\be
&&-i\Big{(}[\varepsilon \stackrel{\ast}{,} \langle\!\langle
\EW^{(n)}_{\mu\nu_1\ldots \nu_{n-1}}\, ,\,h_1^{\nu_1} \ldots h_{n-1}^{\nu_{n-1}}
\rangle\!\rangle]+ {\rm perm}_{n-1} \Big{)}\label{Lnrel2}\\
&&+i \Big{(} \langle\!\langle
\EW^{(n)}_{\mu\nu_1\ldots \nu_{n-1}}\, ,\,[\varepsilon
\stackrel{\ast}{,}h_1^{\nu_1}]h_2^{\nu_2} \ldots
h_{n-1}^{\nu_{n-1}}\rangle\!\rangle+
 \EW^{(n)}_{\mu\nu_1\ldots \nu_{n-1}}\, ,\,[\varepsilon
\stackrel{\ast}{,}h_2^{\nu_1}]h_1^{\nu_2} \ldots
h_{n-1}^{\nu_{n-1}}\rangle\!\rangle\0\\
&&\quad\quad\quad\quad\quad+\ldots + \EW^{(n)}_{\mu\nu_1\ldots \nu_{n-1}}\,
,\,[\varepsilon \stackrel{\ast}{,}h_{n-1}^{\nu_1}]h_1^{\nu_2} \ldots
h_{n-2}^{\nu_{n-1}}\rangle\!\rangle+{\rm perm}_{n-1} \Big{)}\0\\
&& + \frac{1}{n} \Big{(} \langle\!\langle
\EW^{(n+1)}_{\mu\nu_1\ldots \nu_{n}}\, ,\,\partial^{\nu_1}_x\varepsilon\,
h_1^{\nu_2}h_2^{\nu_3} \ldots h_{n-1}^{\nu_{n}}\rangle\!\rangle+{\rm perm}_{n}
\Big{)} =0\0 
\ee
where ${\rm perm}_{n-1}$ means the permutations of $h_1,\ldots , h_{n-1}$, and
${\rm perm}_{n}$ means the permutations
of $h_1,\ldots , h_{n-1}$  {\it and} $\partial_x\varepsilon$.

Now, from \eqref{EFequations} we get 
\be
&&i [\varepsilon \stackrel{\ast}{,}\langle\!\langle \EW_{\mu\nu_1\ldots
\nu_{n-1} }^{(n)}\, ,\,
h^{\mu_1}\ldots h^{\mu_{n-1}}\rangle\!\rangle]    - i \, \sum_{i=1}^{n-1} 
\langle\!\langle \EW_{\mu\mu_1\ldots \mu_i\ldots \mu_{n-1}}^{(n)}\, ,\,
h^{\mu_1}\ldots\,[\varepsilon \stackrel{\ast}{,}h^{\mu_i}]\ldots
h^{\mu_{n-1}}\rangle\!\rangle \0\\
&& -
\frac 1{n}\sum_{i=1}^{n} \langle\!\langle \EW_{\mu\mu_1\ldots
\mu_i\ldots \mu_{n}}^{(n+1)}\,
,\,h^{\mu_1}\ldots\partial_x^{\mu_i}\varepsilon\,  \ldots h^{\mu_{n} }
\rangle\!\rangle=0\label{EFequationsn}
\ee
If now we transform the LHS of this equation to a multilinear function of
$h_1,\ldots, h_{n-1}$ according to the recipe \eqref{symmbydiag}, we obtain
precisely \eqref{Lnrel2}. This completes the proof of the $n$-th $L_\infty$
relation.

\section{Conclusion}

In this paper we have carried out the worldline quantization of a Dirac fermion
field coupled to external sources. In particular, we have determined the formula
for the effective action, by expanding it in
a perturbative series, and determined the generalized equations of motion. This
has allowed us, in the second part of the paper, to show that this set up of the theory
accommodates an $L_\infty$ algebra. We remark that this applies to the full
effective action, i.e. not only to its local part, but also to its non-local
part.  

Although we do not give here an explicit proof, the same symmetry characterizes 
also the effective action obtained by integrating out a scalar field coupled to
the same external sources. The proof in the scalar case is actually easier, because the corresponding
$\EW^{(n)}$'s come out automatically symmetric (for the basic formulas, see \cite {bekaert}).

An $L_\infty$ symmetry is different from the familiar Lie algebra symmetry in
that the equation of motion plays an essential role, in other words the symmetry is dynamical 
(for an early formulation in this sense, see \cite{BBvD}).  The full implications of this (more general)
symmetry are not yet clear. It characterizes a large class 
of {\it perturbative} field theories \cite{HZ}, but certainly not all. For instance, it is not present in the open string field theory  \`a la Witten, where it is replaced by an $A_\infty$ algebra. The classification of field and string theories on the basis of such homotopic-like algebra symmetries is under way. For the time being we intend to use it as a basic working tool in our attempt to generate higher spin theories by integrating out matter fields. 

In what concerns us here, the $L_\infty$ algebra symmetry is a symmetry of the equation of motion of the effective action resulting from integrating out the matter fields. The  $L_\infty$ symmetry descends from the Ward identities of the current correlators of the matter model. 
These Ward identities for current correlators imply the higher spin symmetry of effective action. Therefore one can say that the $L_\infty$  
symmetry is the source of the HS symmetry of the effective action. We recall again that the local part of this action is to be identified, in our approach, with the classical HS action. This has been proved so far only at the quadratic level. That it is true at the interacting levels is the bet of our program. 

Another character of our paper is the worldline quantization. Let us repeat (see introduction) that 
it is not imperative to use the worldline formalism. As we have done in previous papers, one could use the traditional quantization and compute the effective action by means of Feynman diagrams. The problem with this approach is that we do not have a way to fix a priori the form of the currents and the form of the symmetry transformations except by trial and error (a method that becomes rapidly unsustainable for increasing spin). The worldline quantization grants both at the same time.  In this resides the importance of the worldline quantization.

The way we interpret the $L_\infty$ relations among correlators is very similar to the usual Ward identities for an ordinary gauge symmetry 
(we have already pointed out above this parallelism): these relations 
must hold for both the classical and quantum theory, they are the relevant
defining relations.
Their possible breakdown is analogous to the appearances of anomalies in  
ordinary gauge Ward identities. It is interesting that possible obstructions to 
constructing higher spin theories in our scheme might be identified with such
anomalies\footnote{ {It is worth remarking that if such an anomaly occurs
in the WI \eqref{invariance}, i.e. 
$\delta_\varepsilon \EW[h] = \EA[\varepsilon, h]\neq 0$, it must satisfy a consistency
condition, analogous to the WZ condition for the ordinary anomalies: 
$\delta_{\varepsilon_2}\EA[\varepsilon_1, h]-\delta_{\varepsilon_1} \EA[\varepsilon_2, h]=
\EA[[\varepsilon_1\stackrel{\ast}{,} \varepsilon_2],h]$, as a consequence of \eqref{e1e2}}.}.

Finally another consideration: while so far $L_\infty$
algebras have
been discussed mostly in relation to classical (first quantized, in the string
field
theory case) actions, as 
we have remarked above, our $L_\infty$\footnote{In our case we should perhaps
call it $L_\infty^\ast$, 
due to the essential role played in it by the Moyal product.}  structure
characterizes the full effective field action (including its 
non-local part). This is perhaps in keeping with what was noticed in
\cite{BCDGLS,BCDGS}: 
the effective action 
for a single higher spin field, at least at the quadratic order, is
characterized by a unique 
Fronsdal differential operator inflected in various non-local forms.
In any case
it is reassuring
to find such a symmetry in the one-loop effective actions obtained by
integrating out matter fields. Our idea of using this method to generate 
higher spin field theories is perhaps not groundless.


\acknowledgments

{We would like to thank Jim Stasheff for reading the manuscript and for several suggestions.}
This research has been supported by the Croatian Science Foundation under the
project 
No.~8946 and by the University of Rijeka under the research support
No.~13.12.1.4.05. M.\ P.
would like to thank SISSA (Trieste) for support under the Visiting PhD Students
Training Program.

\end{document}